\documentclass[5p]{elsarticle}

\usepackage{hyperref}
\usepackage{graphicx, subfigure}
\usepackage{siunitx}

\begin{document}

\begin{frontmatter}

\title{Preliminary Measurements for a Sub-Femtosecond Electron Bunch Length Diagnostic}


\author[mymainaddress,mysecondaryaddress]{M.K. Weikum\corref{mycorrespondingauthor}}
\cortext[mycorrespondingauthor]{Corresponding Author at: DESY, Bdg. 30b, Notkestr. 85, 22607 Hamburg, Germany. Tel.: +49 40 8998 1584.}
\ead{maria.weikum@desy.de}
\author[mytertiaryaddress,myfourthaddress]{G. Andonian}
\author[myfourthaddress]{N.S. Sudar}
\author[myfifthaddress]{M.G. Fedurin}
\author[myfifthaddress]{M.N. Polyanskiy}
\author[myfifthaddress]{C. Swinson}
\author[mytertiaryaddress]{A. Ovodenko}
\author[mytertiaryaddress]{F. O'Shea}
\author[mytertiaryaddress]{M. Harrison}
\author[mysecondaryaddress]{Z.M. Sheng}
\author[mymainaddress]{R.W. Assmann}

\address[mymainaddress]{Deutsches Elektronensynchrotron (DESY), 22607 Hamburg, Germany}
\address[mysecondaryaddress]{SUPA, Department of Physics, University of Strathclyde, G4 0NG Glasgow, United Kingdom}
\address[mytertiaryaddress]{RadiaBeam Technologies, LLC, Santa Monica, CA 90404, USA}
\address[myfourthaddress]{Department of Physics and Astronomy, UCLA, Los Angeles, CA 90095, USA}
\address[myfifthaddress]{Brookhaven National Laboratory, Long Island, NY 11967, USA}

\begin{abstract}
With electron beam durations down to femtoseconds and sub-femtoseconds achievable in current state-of-the-art accelerators, longitudinal bunch length diagnostics with resolution at the attosecond level are required. In this paper, we present such a novel measurement device which combines a high power laser modulator with an RF deflecting cavity in the orthogonal direction. While the laser applies a strong correlated angular modulation to a beam, the RF deflector ensures the full resolution of this streaking effect across the bunch hence recovering the temporal beam profile with sub-femtosecond resolution. Preliminary measurements to test the key components of this concept were carried out at the Accelerator Test Facility (ATF) at Brookhaven National Laboratory recently, the results of which are presented and discussed here.
Moreover, a possible application of the technique for novel accelerator schemes is examined based on simulations with the particle-tracking code \textit{elegant} and our beam profile reconstruction tool. %Effects limiting the device resolution, in particular the bunch energy spread and initial divergence, are considered in detail in this context.
\end{abstract}

\begin{keyword}
bunch length diagnostic; sub-femtosecond resolution; RF deflecting cavity; laser modulator; ultrashort beam; longitudinal beam profile
\end{keyword}

\end{frontmatter}

%\linenumbers

\section{Introduction}

Most common longitudinal electron beam diagnostic techniques today, such as electro-optical measurements, transition radiation measurements and transverse deflecting cavities, are limited in resolution to the single to tens of femtosecond level \cite{Steffen2009,Heigoldt2015,Nozawa2015,Behrens2014}. This stands in contrast, however, with rapid developments in ultrashort electron beam generation methods aiming at developing bunch lengths in the sub-femtosecond to attosecond range through techniques, like bunch compression \cite{Zhu2016}, microbunching \cite{Sears2008} or laser wakefield acceleration \cite{Tooley2017}. In order to fully characterise such beams longitudinally, current measurement techniques need to be improved or new ones developed, efforts into which in recent years have lead to a range of novel proposals, such as plasma deflectors \cite{Dornmair2016}, THz streaking \cite{Jamison2016} and other phase-space mapping techniques (e.g. \cite{Zhang2017,Huang2010}), most of which have not been experimentally proven so far. \par 
%\begin{figure}[h!]
%	\centering
%	\subfigure{
%		\includegraphics[width=0.143\textwidth]{AttoscopeTheoreticalConcepts-LaserModulator-sxp-afterLMOD-EAACPaper.eps}
%		\label{Theory:LMOD}
%	}
%	\subfigure{
%		\includegraphics[width=0.143\textwidth]{AttoscopeTheoreticalConcepts-Deflector-syp-afterdeflector-EAACPaper.eps}
%		\label{Theory:TDS}
%	}
%	\subfigure{
%		\includegraphics[width=0.143\textwidth]{AttoscopeTheoreticalConcepts-FullAttoscope-EAACPaper.eps}
%		\label{Theory:fullattoscope}
%	}
%	\caption{(a-b): Angular kick experienced by the electrons as a function of their longitudinal coordinate $s$ in the laser modulator (a) and RF deflecting cavity (b). (c): Final transverse beam distribution after the beam has passed through the whole diagnostic device. All three plots are generated based on simulations with the tracking code \textit{elegant} \cite{Borland2000} and visualisation via sirepo \cite{sirepo}.}%, assuming a \SI{5}{\giga\watt} $\textrm{CO}_2$-laser and an RF deflecting voltage of \SI{15}{\mega\volt}, which for (c) is followed by \SI{2}{\metre} drift space.}
%	\label{Theory}
%\end{figure}
\begin{figure}[thb]
	\centering
	\includegraphics[width=0.5\textwidth]{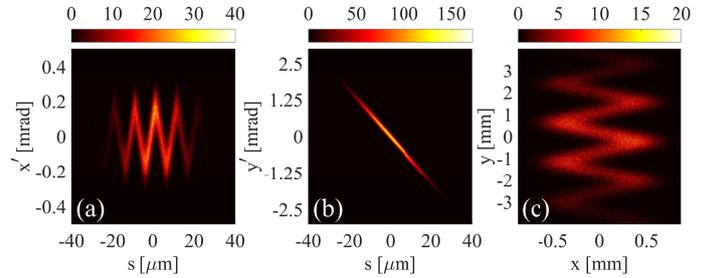}
	\caption{(a-b): Angular kick experienced by the electrons as a function of their longitudinal coordinate $s$ in the laser modulator (a) and RF deflecting cavity (b). (c): Final transverse beam distribution after the beam has passed through both components and a drift space of \SI{2}{\metre}. The color scale shows the electron intensity in arbitrary units. All three plots are generated based on simulations with the tracking code \textit{elegant} \cite{Borland2000} and visualisation via sirepo \cite{sirepo}.}%, assuming a \SI{5}{\giga\watt} $\textrm{CO}_2$-laser and an RF deflecting voltage of \SI{15}{\mega\volt}, which for (c) is followed by \SI{2}{\metre} drift space.}
	\label{Theory}
\end{figure}
\begin{figure*}[th]
	\centering
	\includegraphics[trim=5 90 5 130,clip,width=0.75\textwidth]{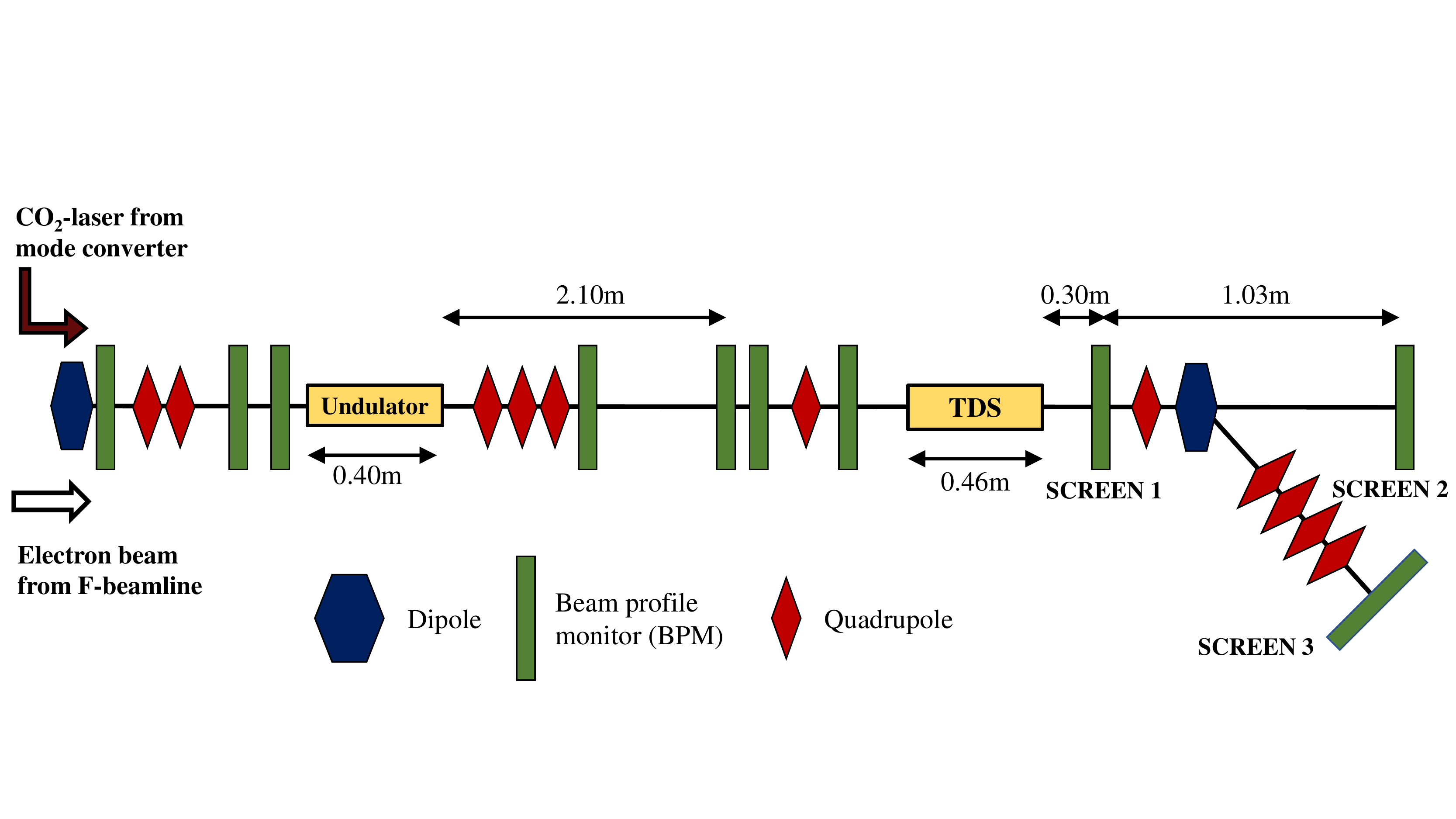}
	\caption{Schematic of the experimental setup in the last section of ATF beamline 2.}
	\label{fig:ExperimentalSetup}
\end{figure*}
In this paper we want to report on experimental tests of one such novel diagnostic device, as proposed by Andonian et al. \cite{Andonian2011}. The machine works by streaking an electron beam both in the horizontal and vertical direction through a combination of a laser modulator -- a laser pulse co-propagating with the electrons in an undulator -- and an RF transverse deflecting structure (TDS). With the laser pulse in the transverse $\textrm{TEM}_{10}$-mode, its interaction with the electron beam in the modulator applies a time-dependent kick to the electrons with amplitude (see Fig.~\ref{Theory}(a))
\begin{equation}
\Delta x^{\prime} = S_{LM} \sin{\left(\omega t+\phi\right)}
\end{equation}
where $\omega$ is the laser frequency and $\phi$ the relative laser-electron phase. $S_{LM}$ is the streaking amplitude of the laser modulator dependent, among others, on laser power, electron energy and undulator parameters. While this deflection can be orders of magnitude stronger than, e.g., from an RF deflecting structure, its sinusoidal dependence with the laser frequency means that the streaking of beam sections in different laser periods overlaps. In order to clearly resolve the full longitudinal beam profile, the electron beam is hence additionally streaked in the orthogonal direction using an RF deflecting cavity to apply a kick of amplitude (see Fig.~\ref{Theory}(b))
\begin{equation}
\Delta y^{\prime} \sim S_{rf}\omega_{rf}t. 
\end{equation}
with $\omega_{rf}$ the RF frequency and $S_{rf}$ the streaking amplitude in y, depending on the RF deflecting voltage and electron energy. This combination allows to separate the signals coming from different beam sections producing a final screen distribution as shown in Fig.~\ref{Theory}(c).
\section{Experimental Setup}
Experimental tests of some of the key concepts of this diagnostic principle were carried out at the Accelerator Test Facility (ATF) at Brookhaven National Laboratory. Figure~\ref{fig:ExperimentalSetup} shows the experimental setup at beamline 2 of ATF with detailed machine parameters listed in Table~\ref{AttoscopeSetupTable}. The depicted section is preceded by an S-band electron gun, linac and beam transport line. A high-power $\textrm{CO}_2$-laser enters the beamline at the first dipole of beamline 2 after it passes through an interferometer stage for mode conversion. Here the interference pattern of two $\textrm{TEM}_{00}$ profiles is generated, providing an accurate representation of the $\textrm{TEM}_{10}$-mode close to the propagation axis. Both laser and electron beam are focused with a long, shallow focus at the centre of the undulator to maximise their interaction, but avoid effects of the laser pulse shape on the electron beam. \par 
One of the most crucial parts in the setup is the temporal synchronisation of the laser pulse and electron beam in the undulator. A rough synchronisation at the picosecond-level was performed by placing a Germanium wafer into the beamline and measuring the laser intensity as the beam-laser delay is scanned based on the electron beam ionising the wafer and consequently blocking the laser light. Further fine-tuning of the delay was achieved by measuring the interaction in the laser modulator as a change in energy spread at the dipole spectrometer (Screen 3). For these measurements, the $\textrm{TEM}_{00}$-mode of the $\textrm{CO}_2$-laser was employed in order to be able to use higher laser power, which is otherwise limited by the damage threshold of components in the interferometer stage as well as energy losses in said section. Based on this fine-timing measurement, the delay between electron beam and laser pulse was adapted to the sub-ps level, which is sufficient for their respective durations. Measurements with the $\textrm{TEM}_{10}$ laser mode were consequently taken with the transverse beam distribution measured at two different diagnostic screens at the end of the beamline, one just \SI{30}{\centi\metre} behind the TDS (Screen 1) and the other about \SI{1.33}{\metre} downstream (Screen~2).

%\begin{itemize}
%	\item figure and table of beamline setup
%	\item procedure for mode conversion and synchronisation
%\end{itemize}

\begin{table}[htb]
	\centering
	\begin{tabular}{ l | r }
		\hline
		\hline
		\textbf{Laser modulator} & \\
		Undulator period, length & \SI{4.0}{\centi\metre}, \SI{0.40}{\metre} \\
		Undulator magnetic field & \SI{0.67}{\tesla} (K=2.5) \\
		Laser wavelength & \SI{10.3}{\micro\metre} \\
		Laser power (in $\textrm{TEM}_{10}$ mode) & \SIrange{2.8}{72}{\giga\watt} \\
		Laser pulse duration & \SI{3.5}{\pico\second} \\
		\hline
		\textbf{RF deflecting cavity} & \\
		RF frequency & \SI{11.424}{\giga\hertz} (X-band) \\
		Deflecting voltage & $\leq$\SI{15}{\mega\volt} \\
		\hline
		\textbf{Electron beam} & \\
		Mean energy & \SI{46}{\mega\electronvolt} \\
		Charge & \SIrange{50}{200}{\pico\coulomb} \\
		Duration & \SIrange{0.05}{2.0}{\pico\second} \\
		\hline
		\hline
	\end{tabular}
	\caption{Parameters of the experimental setup at ATF.}
	\label{AttoscopeSetupTable}
\end{table}

\section{Measurement Results}

\subsection{Synchronisation of electron beam and $\textrm{CO}_2$-laser}
\begin{figure}[h!]
	\centering
	\subfigure{
		\includegraphics[width=0.45\textwidth]{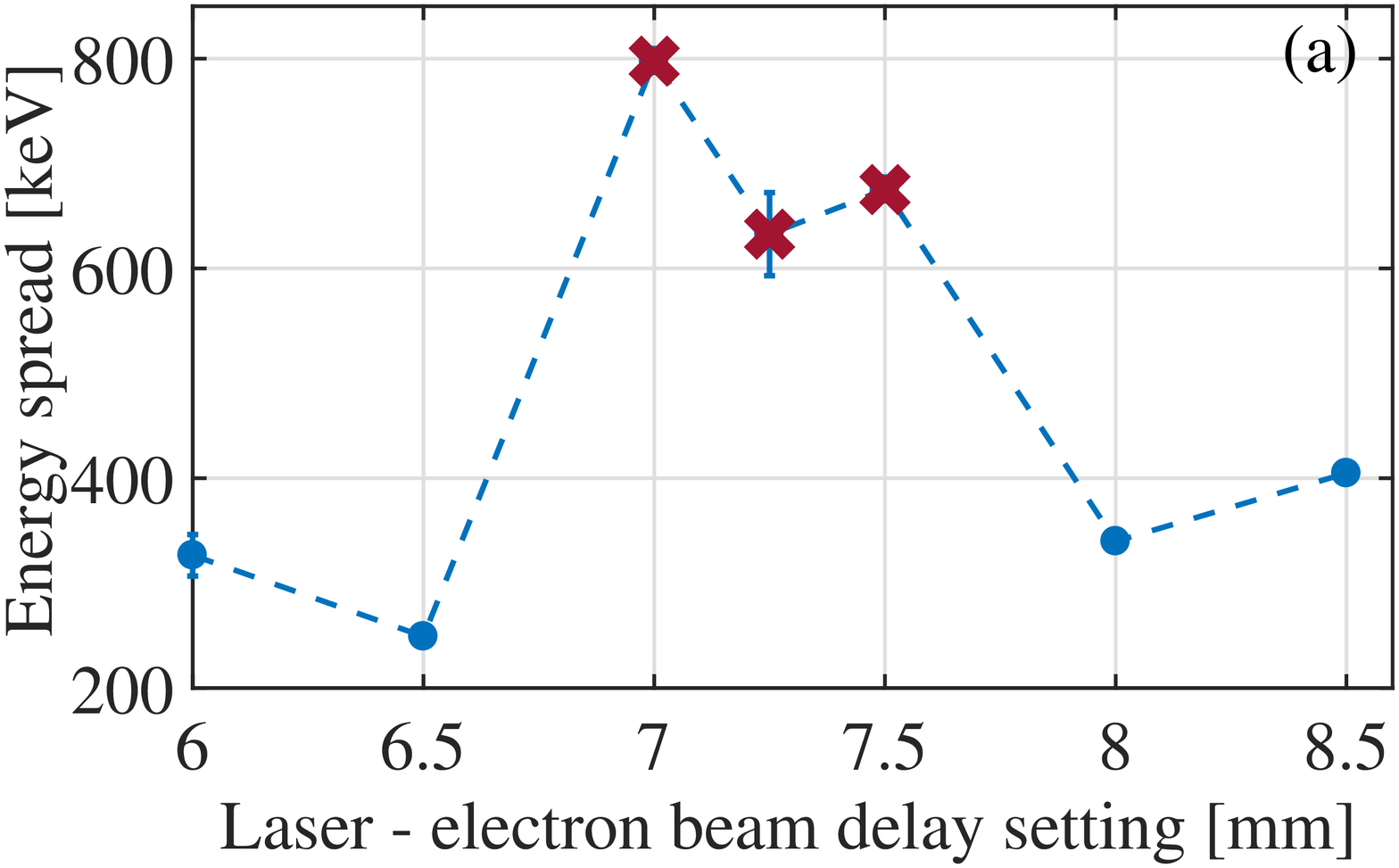}
		\label{fig:synchronisation-delay}
		%\caption{Dependence on delay setting}
	}
	\subfigure{
		\includegraphics[width=0.47\textwidth]{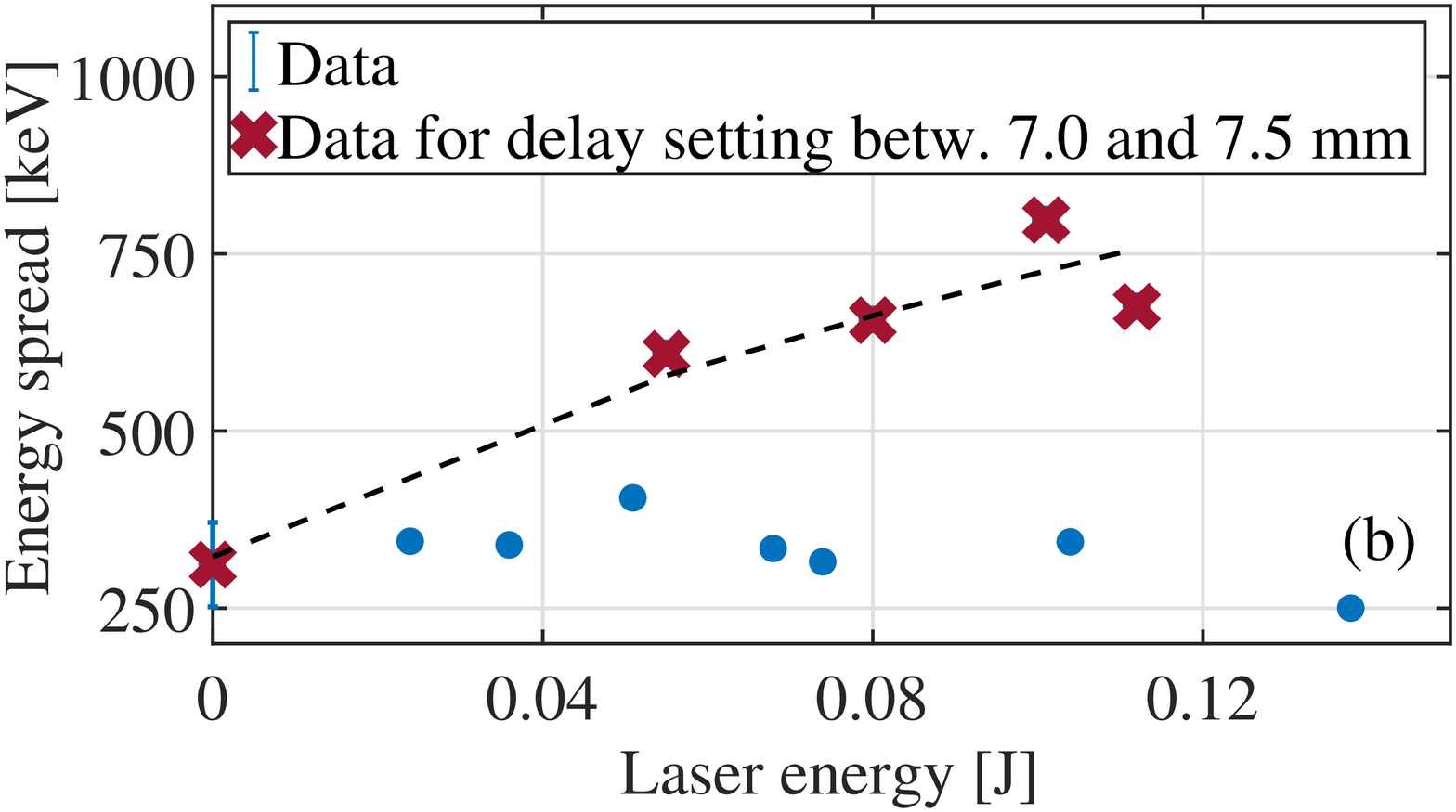}
		\label{fig:synchronisation-energy}
		%\caption{Dependence on laser energy}
	}
	\caption{Fine-tuning of laser-electron beam synchronisation through measuring the beam energy spread due to interaction with the laser pulse in the undulator. The correct synchronisation setting is found at the maximum laser-electron beam interaction for which the energy spread amplitude increases with laser energy. Error bars have been included wherever it was possible to take multiple measurements (between two and four measurements per data point).}
	\label{fig:synchronisation}
\end{figure}

Figure~\ref{fig:synchronisation} shows the results of the fine-tuning of the laser-electron beam delay. The energy spread of the beam, which is broadened due to interaction with the laser pulse in the undulator, is measured at Screen~3 after deflection in the dipole spectrometer. In Fig.~\ref{fig:synchronisation-delay} a clear peak in spread is hence observed for delay settings between \SI{7.0}{\milli\metre} and \SI{7.5}{\milli\metre}. This stage setting corresponds to the ideal overlap between electron beam and laser pulse, such that a maximum interaction between the two occurs. A further confirmation of this delay is provided in Fig.~\ref{fig:synchronisation-energy} where the dependence of the energy spread on applied laser energy is shown. As defined by the crosses in Fig.~\ref{fig:synchronisation}, measurements within the range of ideal settings demonstrate a clear increase of the spread with laser energy. For delay timings outside of this range, however, a change in laser energy has no effect on the measured energy spread, since no strong interaction with the laser pulse takes place. One possible contribution for the shot-to-shot differences observed during synchronisation measurements, as seen from the error bars in Fig.~\ref{fig:synchronisation}, is a timing jitter between laser pulse and electron beam. In order to minimise this effect, however, the duration of the electron beam has been chosen to be small compared to the laser pulse, such that the laser intensity profile shape has little effect on the streaking amplitude. Assuming a constant laser intensity experienced across the electron beam, the streaking measurement itself, as seen in Fig.~\ref{fig:streakedimages}, becomes thus independent of the positioning of the electron beam within the laser pulse, as the jitter simply affects the starting phase from which the beam samples the laser field over multiple periods.

\subsection{Measurements with the $\textrm{TEM}_{10}$-mode}

\begin{figure}[h!]
	\centering
	\subfigure{
		\includegraphics[width=0.43\textwidth]{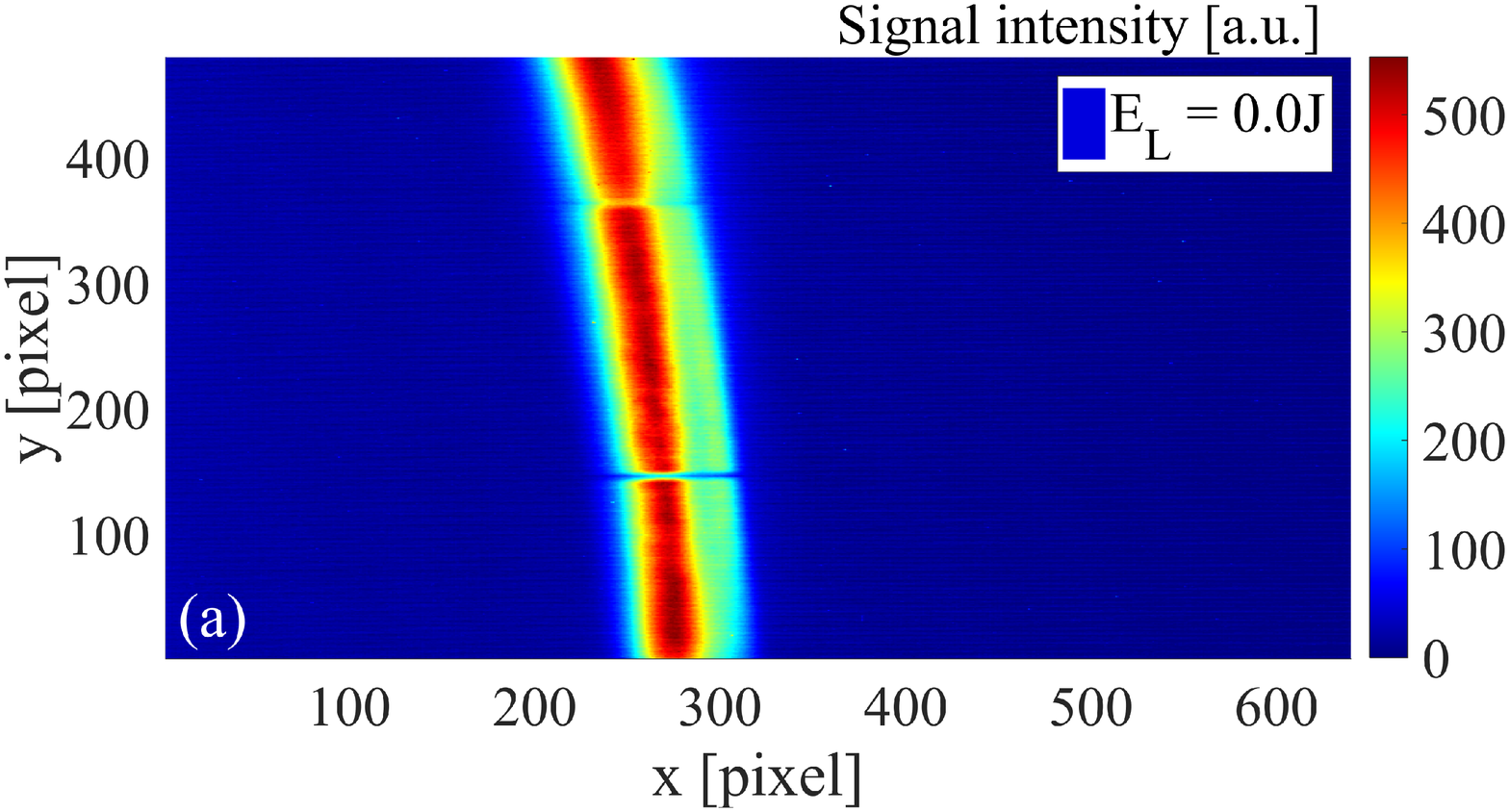}
		\label{fig:EL0_0J}
	}
	\subfigure{
		\includegraphics[width=0.43\textwidth]{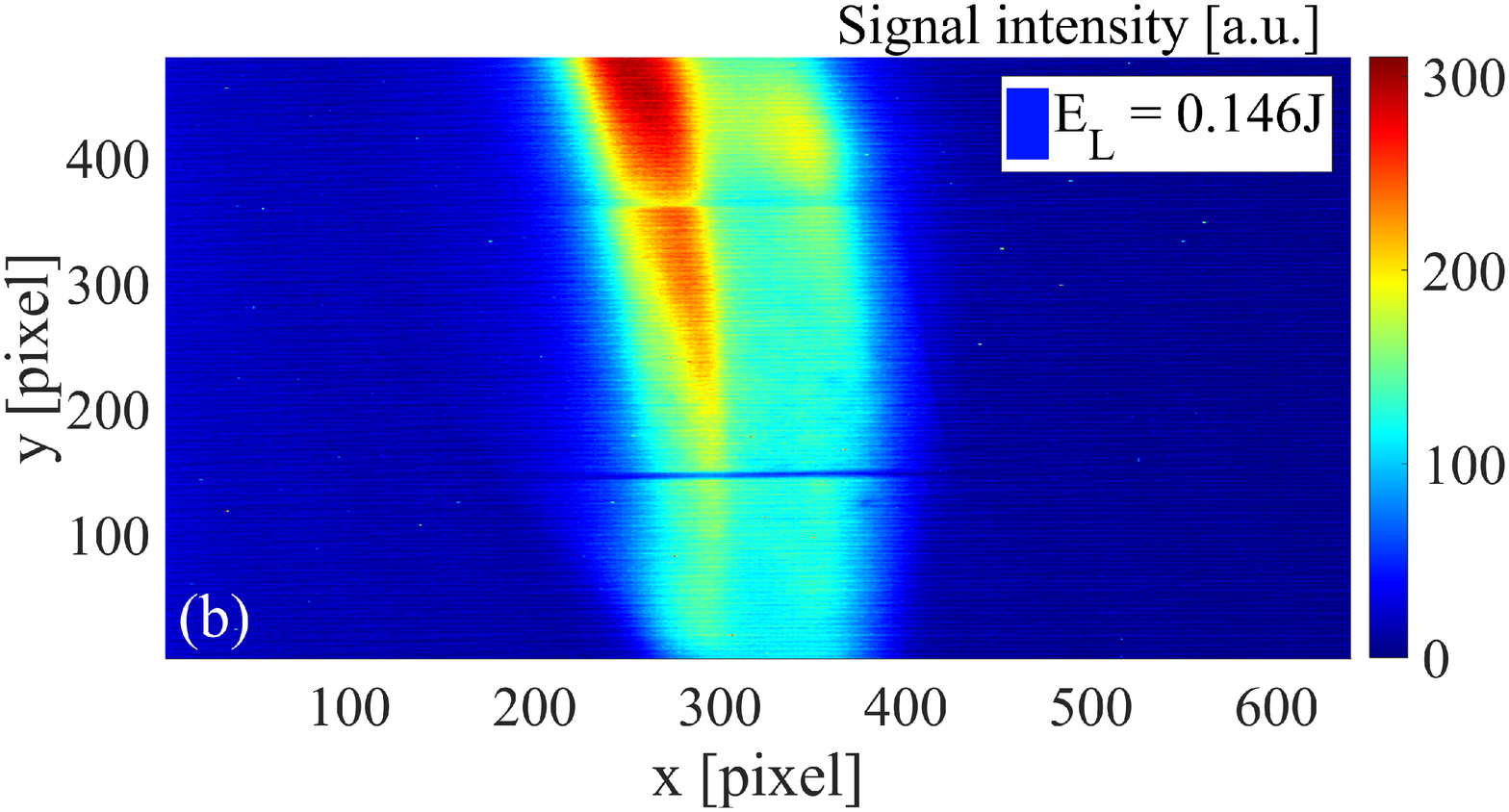}
		\label{fig:EL0_146J}
	}
	\caption{Transverse electron beam distribution at Screen 1 after streaking with the laser modulator and RF deflector. (a) shows the measured screen signal with the laser off, while in (b) it is turned on with a laser energy before mode conversion of $E_L$=\SI{0.146}{\joule}.}
	\label{fig:streakedimages}
\end{figure}

While the interaction of the electron beam with the laser $\textrm{TEM}_{00}$-mode induces an energy spread, it cannot generate a transverse beam spread as is required to use the device for bunch profile diagnostics. For the latter case, hence, the $\textrm{TEM}_{10}$-mode was employed with the same delay setting as determined with the basic laser mode. Figure~\ref{fig:streakedimages} shows the streaking effect of this interaction through two examples of screen images measured at Screen~1 during the experiment. A clear spread in the horizontal beam size is observed, as the $\textrm{CO}_2$-laser is turned on. Additionally, the beam is also stretched in the vertical direction in both images, as a deflecting voltage is applied in the RF TDS. \par 
The typical sinusoidal screen pattern (see Fig.~\ref{Theory}(c)), expected from the combined streaking of RF deflector and laser modulator, was not observable in these first measurements. A qualitative comparison with simulation studies with the tracking code \textit{elegant} identified two main reasons for this behaviour:
\begin{itemize}
\item first, the resolution of the imaging screen is not sufficient to resolve the sinusoidal pattern: assuming a deflecting voltage of \SI[separate-uncertainty = true]{6\pm 3}{\mega\volt} - based on a measurement of the deflector input power and a conservative error estimate of \SI{50}{\percent} - each laser period is resolved over \SI[separate-uncertainty = true]{4.98 \pm 2.49}{{pixels}} vertically on the imaging screen. Considering the large signal intensity range and the finite signal width, this value is not adequate to clearly measure the sinusoidal intensity variations and should be improved in future measurements through, e.g., an increased camera zoom or use of a higher resolution screen.
\item second, the intrinsic, i.e. unstreaked, vertical beam size is dominating the measured screen signal in the vertical direction over the streaking effect. Unlike for the TDS alone, the intrinsic width must be smaller than the streaked vertical size of a single laser period on-screen, as it otherwise smears out the sinusoidal shape of the signal.
\end{itemize}
Both of these issues can easily be mitigated by increasing the vertical streaking with the RF deflecting cavity, which can be more than doubled for the ATF setup. As a consequence, the individual signal turns will be streaked more strongly and become resolvable in the measurement, while the profile of the full beam will no longer fit on the imaging screen. \par 

\section{Application to Ultrashort Electron Beams}
\begin{figure}[htb!]
	\centering
	\includegraphics[width=0.47\textwidth]{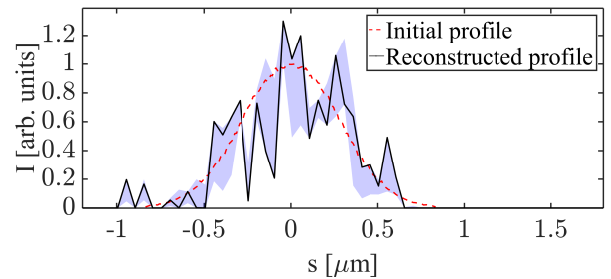}
	\caption{Reconstruction of a short electron beam as possibly achievable from the AXSIS project. Here, a particular working point by T. Vinatier has been tested based on Table~\ref{AXSISSetupTable}. The dashed line shows the original longitudinal beam profile, while the solid line depicts the reconstructed profile and the shaded area is the reconstruction error, estimated based on the uncertainty in the fitting coefficients of the reconstruction algorithm.}
	\label{AXSISBeam-Recon}
\end{figure}

An interesting application for the sub-fs diagnostic beyond the initial demonstration of its operating principle is the measurement of ultrashort electron beams, especially for bunch lengths shorter than half of the laser wavelength where the RF deflector is no longer required. While the machine's applicability to high-quality RF-accelerated electron beams has been discussed in \cite{Weikum2017}, electron bunches from novel accelerators are a more challenging, but also more intriguing regime to explore. One example in this context is the AXSIS (Attosecond X-ray Science: Imaging and Spectroscopy)  project \cite{Kaertner2016} where short bunch duration is expected to be achieved through beam acceleration and compression in THz-driven, dielectric waveguides. Figure~\ref{AXSISBeam-Recon} shows a simulation of the reconstruction of such an example electron beam with the combined diagnostic. Here, the initial bunch properties are based on studies by T. Vinatier for accelerating a beam from an S-band RF gun in a dielectric waveguide \cite{Vinatier2017}. The simulation itself was carried out by tracking the initial electron bunch through the diagnostic setup with the code \textit{elegant} \cite{Borland2000} (with visualisation via the \textit{elegant} implementation of sirepo \cite{sirepo}) and the setup characteristics shown in Table~\ref{AXSISSetupTable}. The reconstruction of the longitudinal beam profile is then completed with a previously developed reconstruction algorithm (see \cite{Weikum2017} for details). \par 
\begin{table}[htb]
	\centering
	\begin{tabular}{ l | r }
		\hline
		\hline
		\textbf{Laser modulator} & \\
		Undulator period, length & \SI{1.5}{\centi\metre}, \SI{4.5}{\centi\metre} \\
		Undulator magnetic field & \SI{0.43}{\tesla} \\
		Laser wavelength & \SI{10.3}{\micro\metre} \\
		Laser power & \SI{100}{\giga\watt} \\
		\hline
		\textbf{Electron beam} & \\
		Mean energy & \SI[separate-uncertainty = true]{15.0\pm 0.24}{\mega\electronvolt} \\
		RMS length & \SI{0.275}{\micro\metre} \\
		Geometric emittance (x,y) & \SI{11}{\nano\metre} \\
		\hline
		\hline
	\end{tabular}
	\caption{Simulation parameters for the AXSIS example beam study. Note that, due to the short bunch length, the RF deflecting cavity behind the laser modulator is optional. Additionally, a Gaussian beam distribution has been assumed, which may need to be refined to a more realistic profile shape in future studies.}
	\label{AXSISSetupTable}
\end{table}
The initial beam profile can be reconstructed well at the sub-femtosecond level with a recovered RMS bunch length of \SI{0.338}{\micro\metre} compared to \SI{0.275}{\micro\metre} as the initial value. The large noise level in the profile intensity is a consequence of the beam collimator with radius \SI{50}{\micro\metre} that is required before the laser modulator to control the transverse beam properties. While the collimator significantly improves the error in the reconstructed bunch length (by a factor of 1.77 compared to a \SI{100}{\micro\metre} radius collimator), it also filters out a large fraction of the beam charge making a measurement of the streaked signal, depending on the full charge, potentially quite challenging. The remaining error in the reconstructed bunch length is dominated by the evolution of the beam inside the diagnostic, where, especially in the undulator - due to the low beam energy and large energy spread - the bunch stretches, as the electron paths become energy dependent. The compactness of the measurement method, additionally to its good resolution, is hence particularly useful in this context to minimise the effect. To further improve the presented results, however, more detailed future studies may centre on the optimisation of the beam focusing and undulator design to reduce both the strong collimation and charge loss as well as the bunch elongation within the diagnostic. % and third, considering the actual longitudinal distribution and phase-space of the beam from the dielectric accelerator, following more advanced studies of the setup.
Additional attention may also need to be placed on the effect of a possible laser-electron beam timing jitter, which, for electron beams covering only a fraction of a laser period, will affect the sampled laser phase region and hence streaking strength and resolution of a measurement, when used as a single-shot diagnostic.

\section{Conclusions and Outlook}
Measurements demonstrating the central components of a combined high-resolution streaking device, using both a laser modulator and RF deflecting cavity, are presented. After fine-tuning of the laser pulse-electron beam synchronisation through a measurement of the beam energy spread caused by the laser interaction, a streaking effect was observed both in the horizontal and vertical direction. A follow-up experimental run at ATF is planned where the focus will lie on resolving the full sinusoidal streaking pattern and better understanding the electron beam dynamics inside the device. Additionally, studies with ultrashort electron beams from novel accelerators are shown to be an interesting potential future application.

\section{Acknowledgements}
This work is supported by the US-DOE Grant No. DE-SC0007701. Additionally, funding has been received from the European Research Council under the European Union’s Seventh Framework Programme (FP/2007-2013) / ERC Grant Agreement N.609920.

\section*{References}

\bibliography{EAAC2017Proceedings-MWeikum-arXiv}

\begin{thebibliography}{10}
\expandafter\ifx\csname url\endcsname\relax
  \def\url#1{\texttt{#1}}\fi
\expandafter\ifx\csname urlprefix\endcsname\relax\def\urlprefix{URL }\fi
\expandafter\ifx\csname href\endcsname\relax
  \def\href#1#2{#2} \def\path#1{#1}\fi

\bibitem{Steffen2009}
B.~Steffen, V.~Arsov, G.~Berden, W.~A. Gillespie, S.~P. Jamison, A.~M. MacLeod,
  A.~F.~G. {van der Meer}, P.~J. Phillips, H.~Schlarb, B.~Schmidt,
  P.~Schm{\"{u}}ser, {Electro-Optic Time Profile Monitors for Femtosecond
  Electron Bunches at the Soft X-Ray Free-Electron Laser FLASH}, Phys. Rev.
  Spec. Top. - Accel. Beams 12 (2009) 032802.
\newblock \href {http://dx.doi.org/10.1103/PhysRevSTAB.12.032802}
  {\path{doi:10.1103/PhysRevSTAB.12.032802}}.

\bibitem{Heigoldt2015}
M.~Heigoldt, A.~Popp, K.~Khrennikov, J.~Wenz, S.~W. Chou, S.~Karsch, S.~I.
  Bajlekov, S.~M. Hooker, B.~Schmidt, {Temporal evolution of longitudinal bunch
  profile in a laser wakefield accelerator}, Phys. Rev. Spec. Top. - Accel.
  Beams 18 (2015) 121302.
\newblock \href {http://dx.doi.org/10.1103/PhysRevSTAB.18.121302}
  {\path{doi:10.1103/PhysRevSTAB.18.121302}}.

\bibitem{Nozawa2015}
I.~Nozawa, M.~Gohdo, K.~Kan, T.~Kondoh, A.~Ogata, J.~Yang, Y.~Yoshida, {Bunch
  Length Measurement of Femtosecond Electron Beam by Monitoring Coherent
  Transition Radiation}, in: Proc. IPAC 2015, Richmond, USA, 2015, pp. 940--943
  (MOPTY002).

\bibitem{Behrens2014}
C.~Behrens, F.-J. Decker, Y.~Ding, V.~A. Dolgashev, J.~Frisch, Z.~Huang,
  P.~Krejcik, H.~Loos, A.~Lutman, T.~J. Maxwell, J.~Turner, J.~Wang, M.-H.
  Wang, J.~Welch, J.~Wu, {Few-femtosecond time-resolved measurements of X-ray
  free-electron lasers}, Nat. Commun. 5 (2014) 3762.
\newblock \href {http://dx.doi.org/10.1038/ncomms4762}
  {\path{doi:10.1038/ncomms4762}}.

\bibitem{Zhu2016}
J.~Zhu, R.~Assmann, U.~Dorda, B.~Marchetti, {Matching sub-fs electron bunches
  for laser-driven plasma acceleration at SINBAD}, Nucl. Instruments Methods
  Phys. Res. Sect. A Accel. Spectrometers, Detect. Assoc. Equip. 829 (2016)
  229--232.
\newblock \href {http://dx.doi.org/10.1016/j.nima.2016.01.066}
  {\path{doi:10.1016/j.nima.2016.01.066}}.

\bibitem{Sears2008}
C.~M.~S. Sears, E.~Colby, R.~Ischebeck, C.~McGuinness, J.~Nelson, R.~Noble,
  R.~H. Siemann, J.~Spencer, D.~Walz, T.~Plettner, R.~L. Byer, {Production and
  characterization of attosecond electron bunch trains}, Phys. Rev. Spec. Top.
  - Accel. Beams 11 (2008) 061301.
\newblock \href {http://dx.doi.org/10.1103/PhysRevSTAB.11.061301}
  {\path{doi:10.1103/PhysRevSTAB.11.061301}}.

\bibitem{Tooley2017}
M.~P. Tooley, B.~Ersfeld, S.~R. Yoffe, A.~Noble, E.~Brunetti, Z.~M. Sheng,
  M.~R. Islam, D.~A. Jaroszynski, {Towards Attosecond High-Energy Electron
  Bunches : Controlling Self-Injection in Laser Wakefield Accelerators through
  Plasma Density Modulation}, Phys. Rev. Lett. 119 (2017) 044801.
\newblock \href {http://dx.doi.org/10.1103/PhysRevLett.119.044801}
  {\path{doi:10.1103/PhysRevLett.119.044801}}.

\bibitem{Dornmair2016}
I.~Dornmair, C.~B. Schroeder, K.~Floettmann, B.~Marchetti, A.~R. Maier,
  {Plasma-driven ultrashort bunch diagnostics}, Phys. Rev. Accel. Beams 19
  (2016) 062801.
\newblock \href {http://dx.doi.org/10.1103/PhysRevAccelBeams.19.062801}
  {\path{doi:10.1103/PhysRevAccelBeams.19.062801}}.

\bibitem{Jamison2016}
S.~P. Jamison, E.~W. Snedden, D.~A. Walsh, M.~J. Cliffe, D.~M. Graham, D.~S.
  Lake, {A THz Driven Transverse Deflector for Femtosecond Longitudinal Profile
  Diagnostics}, in: Proc. IBIC 2016, Barcelona, Spain, 2016, pp. 748--751
  (WEPG48).

\bibitem{Zhang2017}
Z.~Zhang, Y.~Du, C.~Tang, Y.~Ding, Z.~Huang, {Optical Circular Deflector with
  Attosecond Resolution for Ultrashort Electron Beam}, Phys. Rev. Accel. Beams
  20 (2017) 050702.
\newblock \href {http://dx.doi.org/10.1103/PhysRevAccelBeams.20.050702}
  {\path{doi:10.1103/PhysRevAccelBeams.20.050702}}.

\bibitem{Huang2010}
Z.~Huang, K.~Bane, Y.~Ding, P.~Emma, {Single-shot method for measuring
  femtosecond bunch length in linac-based free-electron lasers}, Phys. Rev.
  Spec. Top. - Accel. Beams 13 (2010) 092801.
\newblock \href {http://dx.doi.org/10.1103/PhysRevSTAB.13.092801}
  {\path{doi:10.1103/PhysRevSTAB.13.092801}}.

\bibitem{Borland2000}
M.~Borland, {Elegant: A flexible SDDS-compliant code for accelerator
  simulation}, Adv. Phot. Source LS-287 (2000) 1--11 .\href
  {http://dx.doi.org/10.2172/761286} {\path{doi:10.2172/761286}}.

\bibitem{sirepo}
{Radiasoft}, \href{https://beta.sirepo.com}{{Sirepo}}, (accessed on 14/11/2017)
  (2017).
\newline\urlprefix\url{https://beta.sirepo.com}

\bibitem{Andonian2011}
G.~Andonian, E.~Hemsing, D.~Xiang, P.~Musumeci, A.~Murokh, S.~Tochitsky, J.~B.
  Rosenzweig, {Longitudinal Profile Diagnostic Scheme with Subfemtosecond
  Resolution for High-Brightness Electron Beams}, Phys. Rev. Spec. Top. -
  Accel. Beams 14 (2011) 072802.
\newblock \href {http://dx.doi.org/10.1103/PhysRevSTAB.14.072802}
  {\path{doi:10.1103/PhysRevSTAB.14.072802}}.

\bibitem{Weikum2017}
M.~K. Weikum, G.~Andonian, R.~W. Assmann, U.~Dorda, Z.~M. Sheng,
  {Reconstruction of sub-femtosecond longitudinal bunch profile measurement
  data}, J. Phys. Conf. Ser. 874~(1) (2017) 012079.
\newblock \href {http://dx.doi.org/10.1088/1742-6596/874/1/012079}
  {\path{doi:10.1088/1742-6596/874/1/012079}}.

\bibitem{Kaertner2016}
F.~X. K{\"{a}}rtner, F.~Ahr, A.-L. Calendron, H.~Cankaya, S.~Carbajo, G.~Chang,
  G.~Cirmi, K.~D{\"{o}}rner, U.~Dorda, A.~Fallahi, A.~Hartin, M.~Hemmer,
  R.~Hobbs, Y.~Hua, W.~R. Huang, R.~Letrun, N.~Matlis, V.~Mazalova, O.~D.
  M{\"{u}}cke, E.~Nanni, W.~Putnam, K.~Ravi, F.~Reichert, I.~Sarrou, X.~Wu,
  A.~Yahaghi, H.~Ye, L.~Zapata, D.~Zhang, C.~Zhou, R.~J.~D. Miller, K.~K.
  Berggren, H.~Graafsma, A.~Meents, R.~W. Assmann, H.~N. Chapman, P.~Fromme,
  {AXSIS: Exploring the Frontiers in Attosecond X-Ray Science, Imaging and
  Spectroscopy}, Nucl. Instruments Methods Phys. Res. Sect. A Accel.
  Spectrometers, Detect. Assoc. Equip. 829 (2016) 24--29.
\newblock \href {http://dx.doi.org/10.1016/j.nima.2016.02.080}
  {\path{doi:10.1016/j.nima.2016.02.080}}.

\bibitem{Vinatier2017}
T.~Vinatier, R.~Assmann, U.~Dorda, F.~Lemery, B.~Marchetti, Simulations on a
  potential hybrid and compact attosecond x-ray source based on rf and thz
  technologies, (at European Advanced Accelerator Concepts (EAAC) Workshop
  2017, Elba, Italy). \url{
  https://agenda.infn.it/materialDisplay.py?contribId=169&sessionId=15&materialId=slides&confId=12611}
  (2017).

\end{thebibliography}

\end{document}